\documentclass[journal,onecolumn,12pt,twoside]{IEEEtran}
\usepackage{setspace}
\usepackage{subfig}
\usepackage{caption}
\usepackage{float}
\usepackage{setspace}
\usepackage{graphicx}
\usepackage{graphicx,dblfloatfix}
\usepackage{tikz}
\usepackage{color}
\usepackage{amsmath}
\usepackage{amssymb} 

\usetikzlibrary{shapes,arrows,shadows}
\usepackage{xcolor}
\usepackage{lipsum}

\ifCLASSINFOpdf
\else
\fi
\usepackage{capt-of}

\usepackage{caption}
\ifCLASSINFOpdf
\else
\fi
\begin{document}
%
\title{Intra-Channel Nonlinearity Compensation Based on Second-Order Perturbation Theory}
%
%
%

\author{O.~S.~Sunish Kumar,~\IEEEmembership{Student Member,~IEEE,}
	A.~Amari,~\IEEEmembership{Member,~IEEE}, O.~A.~Dobre,~\IEEEmembership{Fellow,~IEEE}, and~R.~Venkatesan,~\IEEEmembership{Life Senior Member,~IEEE}
	\thanks{O. S. Sunish Kumar, O. A. Dobre, and R. Venkatesan are with the Department
		of Computer Engineering,  Memorial University, St. John’s, NL, A1B 3X5, Canada, e-mail: skos71@mun.ca.}
	\thanks{A. Amari is with the VPIphotonics GmbH, Carnotstr. 6, 10587 Berlin, Germany.}}

\maketitle

\begin{abstract}
The first-order (FO) perturbation theory has been widely investigated to design the digital nonlinearity compensation (NLC) technique to deal with the intra-channel fiber nonlinearity effect in coherent optical communication systems. The main advantages of the perturbation theory-based approach are the possibility of the implementation on a single stage for the entire fiber link and one sample per symbol operation. In this paper, we propose to extend the FO perturbation theory-based NLC (FO-PB-NLC) technique to the second-order (SO), referred to as the SO-PB-NLC, to enhance the NLC performance. We present a comprehensive theoretical analysis for the derivation of the SO nonlinear distortion field, which is the foundation for the SO-PB-NLC technique. Through numerical simulations, we show that the proposed SO-PB-NLC technique significantly enhances the NLC performance and the maximum transmission reach when compared to the FO-PB-NLC technique. Then, the performance of the SO-PB-NLC technique is compared with that of the benchmark digital back-propagation (DBP).       
\end{abstract}

%
\section{Introduction}
%
%
%
%
\IEEEPARstart{I}{n} recent years, the increased usage of bandwidth-intensive applications such as virtual reality and cloud services, as well as Internet-of-Things dramatically increased the network traffic in the core communication network \cite{MChen2019}-\cite{FRestuccia2018}. That necessitates the development of high data-rate optical communication systems to handle such traffic surges. The modern high data-rate optical transmission systems use multilevel modulation formats, which require a higher optical signal-to-noise ratio (OSNR). However, the optical intensity-dependent nonlinear Kerr effect significantly degrades the transmission performance as the fiber launch power increases \cite{Malekiha2016}, \cite{Agrawal1995}. In a dispersion unmanaged optical communication system, the signal-to-signal intra-channel Kerr nonlinearity is considered a dominant impairment, which limits the transmission performance \cite{Malekiha2016}. It is worth mentioning that the signal-to-signal intra-channel nonlinearity can be compensated in principle due to its deterministic nature \cite{Malekiha2016}.

It was shown a few years ago that digital compensation of the intra-channel fiber nonlinearity impairment can be achieved using coherent detection and digital signal processing. Digital back-propagation (DBP) is an extensively investigated fiber nonlinearity compensation (NLC) technique, which uses the numerical solution of the nonlinear Schrödinger equation (NLSE) \cite{Irukulapati2014}. The DBP technique can jointly compensate for the chromatic dispersion (CD) and fiber nonlinearity using the split-step Fourier method (SSFM) with appropriate inverted channel parameters. SSFM is an iterative method which splits the fiber span into several slices, each containing a linear and a nonlinear computation stage \cite{CAHall1990}. The accuracy of the DBP technique can be improved by increasing the number of iterations used in the SSFM algorithm with a corresponding increase in the computational complexity \cite{CSMartins2018}-\cite{FZhang2016}. Owing to its high accuracy, the DBP technique is often used to benchmark other NLC techniques proposed in the literature. 
 
In contrast to SSFM, solutions of the NLSE can be analytically approximated using Volterra series, which is a well-established tool in nonlinear systems theory \cite{Guiomar2011}. In a Volterra series-based approach, the input-output relationship of a nonlinear fiber channel can be represented by a series of nonlinear kernel functions, referred to as Volterra series transfer functions (VSTFs) \cite{AAmariICTON2017}. The first step in the Volterra series-based approach is to model the optical fiber channel based on VSTF \cite{VgenopoulouACPC2014}, \cite{ABakhshali2016}. Then, the \textit{p}-th order theory proposed in \cite{Schetzen1980} is used to derive the inverse-VSTF kernels as a function of the VSTF ones. These kernels are used to design a nonlinear equalizer that compensates for the CD and fiber nonlinearity \cite{Diamantopoulos2019}. 

In contrast to the Volterra series-based approach, the NLSE can be analytically solved using the perturbation theory \cite{Mecozzi2000}. In this approach, the solution of NLSE can be expanded as an infinite power series of the fiber nonlinearity coefficient. Such an iterative method provides a closed-form approximation of the nonlinearly distorted signal field, which imparts a good insight into the nature of the interaction between CD and Kerr nonlinearity. The first-order (FO) perturbation theory (the perturbation series approximation truncated to FO) was initially used to model the intra-channel nonlinearity distortion between highly dispersive and ultra-short Gaussian pulses propagating in the optical fiber link \cite{Mecozzi2000}. These results were later adopted in to design an FO perturbation theory-based NLC\footnote{In the perturbation theory-based approach, NLC is often referred to as either predistortion or post-compensation method. In our work on the perturbation theory-based technique, NLC refers to the predistortion method to compensate for the fiber nonlinearity throughout the paper.} (FO-PB-NLC) to deal with the detrimental effects of fiber nonlinearity. The central idea of the FO-PB-NLC technique is to consider nonlinear distortion as a perturbation correction to the unperturbed solution of the NLSE \cite{Silva2019}-\cite{Dou2012}. It is important to mention that the unperturbed solution takes into account only the distortion due to the CD with the assumption that nonlinearity is absent. The main advantage of the perturbation theory-based approach is the possibility of a single-stage implementation for the entire fiber link \cite{Oyama2014}. It also facilitates one sample per symbol processing, which relaxes the stringent requirement on the electronic hardware speed \cite{Oyama2014}. However, the NLC performance decreases as the launch power increases for the FO-PB-NLC technique. This is attributed to the fact that the FO perturbation series approximation becomes inaccurate to model the nonlinear phase shift as the launch power increases. The use of higher-order modulation formats in the high data-rate optical communication system increases the transmit launch power, and thereby, the higher-order perturbation terms become significant in such transmission systems. In this paper, we propose the extension of the FO-PB-NLC technique to the second-order (SO), referred to as the SO-PB-NLC, to improve the NLC performance.  
    
 \section{System Model}
 \begin{figure*}[!t]
 	\begin{centering}
 		\includegraphics[width=0.75\paperwidth,height=0.16\paperheight]{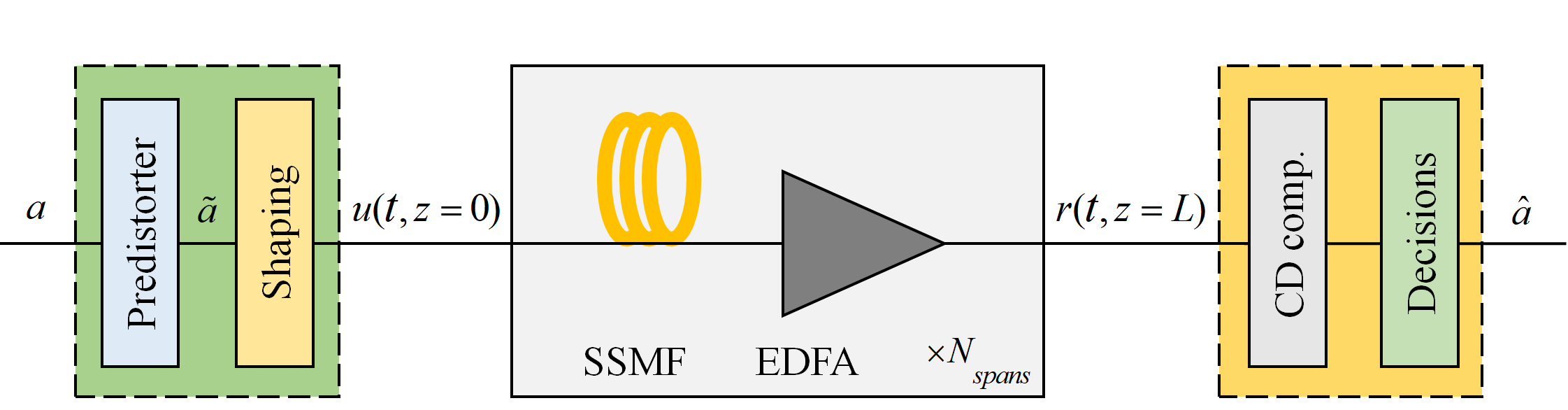}
 		\par\end{centering}
 	\caption{The system model.}
 	\rule[0.5ex]{1\textwidth}{0.5pt}
 \end{figure*}
                  
\subsection{High-Level Description}
The system model, shown in Fig. 1,  comprises a perturbation theory-based predistorter and a pulse shaper at the transmitter, a fiber-optic transmission link with $N_{spans}$ spans of standard single-mode fiber (SSMF), and the receiver consisting of a CD post-compensator followed by a decision unit. In each fiber span, an erbium-doped fiber amplifier (EDFA) is employed for the periodic amplification of the optical signal to compensate for the fiber attenuation. 

A sequence of $K$ symbols $\textbf{a}=[a_{1},a_{2},...,a_{K}]\in\Omega^{K},$ with $\Omega$ is the symbol alphabet, is predistorted first, and then, input to a shaping filter $g(\acute{t}),$ where $\acute{t}$ as the time variable. The resultant signal can be represented as $u(\acute{t},z=0)=\sum_{k=1}^{K}a_{k}g(\acute{t}-kT),$ where $z$ is the space variable and $T$ is the symbol duration. After pulse shaping, the signal is up-converted to the optical domain and transmitted over the fiber-optic transmission link. At the receiver, after down-conversion to the electrical domain, the baseband signal field can be represented as $r(\acute{t},z=L),$ where $L$ is the transmission length. Then, the accumulated\footnote{In a typical dispersion unmanaged optical transmission system, the accumulated CD is compensated in electrical domain using frequency-domain equalizers employing overlap-add/overlap-save algorithm \cite{YLi2008}.} CD is compensated in the electrical domain. Finally, we employ a symbol-by-symbol maximum likelihood detection to carry out the symbol decisions. 
      
\subsection{Signal Propagation Model} 
In this subsection, we describe the model of signal propagation in the optical fiber channel. In a single-mode optical fiber channel, the propagation of the optical field complex envelop $q(z,\acute{t})$ can be modeled by using the NLSE (noiseless) as:

\begin{equation}
	\label{eqn1}
	\frac{\partial}{\partial z}q(z,\acute{t})+\frac{\alpha}{2}q(z,\acute{t})+j\frac{\beta_{2}}{2}\frac{\partial^{2}}{\partial\acute{t}^{2}}q(z,\acute{t})=j\gamma\left|q(z,\acute{t})\right|^{2}q(z,\acute{t}),
\end{equation}
where $\alpha$ is the attenuation, $\beta_{2}$ is the group velocity dispersion, $\gamma$ is the nonlinearity coefficient, and $z$ is the transmission distance.
The NLSE can be further simplified by introducing a normalized field $u(z,t)$ referred to the delayed time frame $t=\acute{t}-(z/v_{g})$ corresponding to the group velocity $v_{g}.$ Thus, by applying the transformation $q(z,\acute{t})\triangleq u(z,t)\exp(-\frac{\alpha}{2}z),$ (\ref{eqn1}) can be modified as: 

\begin{equation}
\label{eqn2}
\frac{\partial}{\partial z}u(z,t)+j\frac{\beta_{2}}{2}\frac{\partial^{2}}{\partial t^{2}}u(z,t)=j\gamma\left|u(z,t)\right|^{2}u(z,t)\exp(-\alpha z).
\end{equation}

The propagation model in (\ref{eqn2}) can be rearranged by separating the linear and nonlinear parts as:

\begin{align}
\label{eqn3}
\frac{\partial}{\partial z}u(z,t) & =\left(\hat{D}+\hat{N}\right)u(z,t)\\
\label{eqn4}
\hat{D} & =-j\frac{\beta_{2}}{2}\frac{\partial^{2}}{\partial t^{2}}\\
\label{eqn5}
\hat{N} & =j\gamma\left|u(z,t)\right|^{2}\exp(-\alpha z),
\end{align}
where $\hat{D}$ and $\hat{N}$ are the linear and nonlinear operators.

The NLSE in (\ref{eqn3}) cannot be solved analytically, except for some special cases \cite{DZwillinger1989}. Numerical approaches, such as SSFM, is typically used to solve the propagation equation in (\ref{eqn3}). In the SSFM, the nonlinear and dispersive signal propagation in the optical fiber is iteratively modeled by dividing the fiber spans into small segments, each having a length of $h$. The step size $h$ is chosen small enough such that the nonlinear and linear effects in each segment can be modeled as acting independently. The symmetric SSFM can be represented as follows:

\begin{equation}
\label{eqn6}
u(z+h,t)=\exp\left(\frac{h}{2}\hat{D}\right)\exp\left(\intop_{z}^{z+h}\hat{N}(z')dz'\right)
\exp\left(\frac{h}{2}\hat{D}\right)u(z,t).
\end{equation}   
   
The accuracy of the SSFM can be improved by increasing the number of iterations and by decreasing the step size $h$. In this study, we use SSFM to model the evolution of the optical field envelops in the optical fiber channel. 
   
\section{Perturbation Theory-Based NLC: Preliminaries}

The direct numerical solution of the NLSE using the SSFM was initially adopted as the key design tool by the optical communication community \cite{CAHall1990}. However, the implementation complexity of the SSFM-based nonlinear fiber propagation modeling was considered impractically high. That led to increased interest in research for the simplified versions of the NLSE for which an approximate analytical solution is available. In contrast to SSFM, the regular perturbation (RP) theory-based method provides an approximate analytical solution of the NLSE in a computationally efficient way. The RP method provides a recursive closed-form solution for the NLSE that gives a good insight into the nature of the interaction between the accumulated CD and the Kerr nonlinearity in the optical fiber channel.

In the RP method, the optical field $u(z,t)$ is expressed in a power series of the nonlinearity coefficient $\gamma$ as $u(z,t)=\sum_{k'=0}^{\infty}\gamma^{k'}u_{k'}(z,t)$, where $k'$ is the order of the perturbative solution. Then, substituting $u(z,t)$ in (\ref{eqn2}), we get \cite{Mecozzi2000}:

\begin{multline}
\label{eqn7}
\sum_{k'=0}^{\infty}\gamma^{k'}\frac{\partial}{\partial z}u_{k'}(z,t)=-\sum_{k'=0}^{\infty}\gamma^{k'}j\frac{\beta_{2}}{2}\frac{\partial^{2}}{\partial t^{2}}u_{k'}(z,t)\\
+j\gamma\sum_{m=0}^{\infty}\sum_{l=0}^{\infty}\sum_{n=0}^{\infty}\gamma^{m+l+n} u_{m}(z,t)u_{l}^{*}(z,t)u_{n}(z,t)\exp(-\alpha z).
\end{multline}

From (\ref{eqn7}), a system of recursive linear differential equations is obtained by equating the terms that multiply equal powers of $\gamma$ on both sides of the equal sign. The differential equation governing the $k'^{th}$- order solution can be represented as:

\begin{equation}
\label{eqn8}
\frac{\partial}{\partial z}u_{k'}(z,t)=-j\frac{\beta_{2}}{2}\frac{\partial^{2}}{\partial t^{2}}u_{k'}(z,t)+j\sum\sum_{m+l+n=k'-1}\sum
u_{m}(z,t)u_{l}^{*}(z,t)u_{n}(z,t)\exp(-\alpha z).
\end{equation}

\subsection{Zeroth-Order (or Linear) Solution} 
The differential equation governing the zeroth-order (or linear) solution is obtained by substituting $k'=0$ in (\ref{eqn8}), which can be represented as \cite{Mecozzi2000}:

\begin{equation}
\label{eqn9}
\frac{\partial}{\partial z}u_{0}(z,t)=-j\frac{\beta_{2}}{2}\frac{\partial^{2}}{\partial t^{2}}u_{0}(z,t).
\end{equation}

By solving (\ref{eqn9}), the zeroth-order solution at a transmission length $z=L$ is obtained as:

\begin{equation}
\label{eqn10}
u_{0}(L,t)=u(0,t)\otimes h_{L}(t),
\end{equation}
where $\otimes$ is the convolution operation, $h_{L}(t)=\mathcal{F}^{-1}\{\exp(-j\frac{w^{2}\beta_{2}L}{2})\}=\frac{1}{\sqrt{-2\pi j\beta_{2}z}}\exp\left(\frac{-jt^{2}}{2\beta_{2}z}\right)$ at the angular frequency $w,$ and $\mathcal{F}^{-1}\{.\}$ is the inverse Fourier transform operation.   

\subsection{FO Solution}    
By substituting $k'=1$ in (\ref{eqn8}), the differential equation governing the FO solution can be represented as:

\begin{equation}
\label{eqn11}
\frac{\partial}{\partial z}u_{1}(z,t)=-j\frac{\beta_{2}}{2}\frac{\partial^{2}}{\partial t^{2}}u_{1}(z,t)+j\left|u_{0}(z,t)\right|^{2}u_{0}(z,t)\exp(-\alpha z).
\end{equation}

The FO distortion field at a transmission distance $z=L$ is obtained by solving (\ref{eqn11}), which can be represented as \cite{Mecozzi2000}:

\begin{equation}
\label{eqn12}
u_{1}(L,t)=\intop_{0}^{L}\exp(-j\frac{w^{2}\beta_{2}z}{2})\exp(-\alpha z)\left(h_{z}(t)\vphantom{\otimes[\left|u_{0}(z,t)\right|^{2}u_{0}(z,t)]}\right.\left.\otimes[\left|u_{0}(z,t)\right|^{2}u_{0}(z,t)]\right)dz\exp(j\frac{w^{2}\beta_{2}L}{2}).
\end{equation}

\subsection{FO-PB-NLC Technique} 
The FO-PB-NLC technique relies on some simplifying assumptions in deriving the approximate FO nonlinear distortion field using (\ref{eqn12}), including \cite{Mecozzi2000}:

\begin{itemize}
	\item The accumulated CD is fully compensated electronically at the receiver.
	\item The input pulses are Gaussian shaped.   
\end{itemize}

Based on the FO perturbation theory, three input Gaussian pulses $\sqrt{P_{0}}a_{m/l/n}\exp(-(t-T_{m/n/l})^{2}/2\tau^{2}),$
at three time instants $T_{m},\,T_{l},\,T_{n}$ generate a ghost pulse due to the nonlinear interaction as follows \cite{Mecozzi2000}:

\begin{multline}
\label{eqn13}
u_{1}(L,t+kT)=j\gamma P_{0}^{3/2}\sum_{m}\sum_{l}\sum_{n}a_{m}a_{l}^{*}a_{n}\exp(\frac{-t^{2}}{6\tau^{2}})\intop_{0}^{L}\frac{\exp(-\alpha z)}{\sqrt{1+2j\beta_{2}z/\tau^{2}+3(\beta_{2}z/\tau^{2})^{2}}}\\
\hspace{-2.2cm}\times\exp\left\{ \begin{array}{c}
-\frac{3\left[2t/3+(m-n)T\right]\left[2t/3+(n-l)T\right]}{\tau^{2}(1+3j\beta_{2}z/\tau^{2})}\\
-\frac{(n-m)^{2}T^{2}}{\tau^{2}\left[1+2j\beta_{2}z/\tau^{2}+3(\beta_{2}z/\tau^{2})^{2}\right]}
\end{array}\right\} dz,
\end{multline}
where $k=m+n-l$, $m,\,n,\,l$ are the symbol indices, $P_{0}$ is the peak power, $a_{m/l/n}$ is the symbol complex amplitude, and $\tau$ is the pulse width.

Fig. 2(a) shows a schematic representation of the triplet pulses involved in the calculation of the FO nonlinear distortion field using (\ref{eqn13}). Without loss of generality, in the predistortion technique, the perturbation of the symbol at index $k=0$, i.e., $l=m+n$ is calculated. The predistortion is assumed to operate at the symbol rate; therefore, the perturbation value at $t=0$ is calculated. Accordingly, (\ref{eqn13}) can be further simplified as:

\begin{equation}
\label{eqn14}
u_{1}(L,t)=j\gamma P_{0}^{3/2}\sum_{m}\sum_{n}a_{m}a_{m+n}^{*}a_{n}\textbf{C}_{m,n}^{FO},
\end{equation}
where $*$ represents the complex conjugate operation and $\textbf{C}_{m,n}^{FO}$ is the FO perturbation coefficient matrix, which is given as:
\begin{figure*}[t]
	\noindent\begin{minipage}[t]{0.6\columnwidth}%
		\begin{center}
			\hspace{-4.5cm}%
			\begin{minipage}[t]{0.48\columnwidth}%
				\begin{flushleft}
					\includegraphics[width=0.28\paperwidth,height=0.09\paperheight]{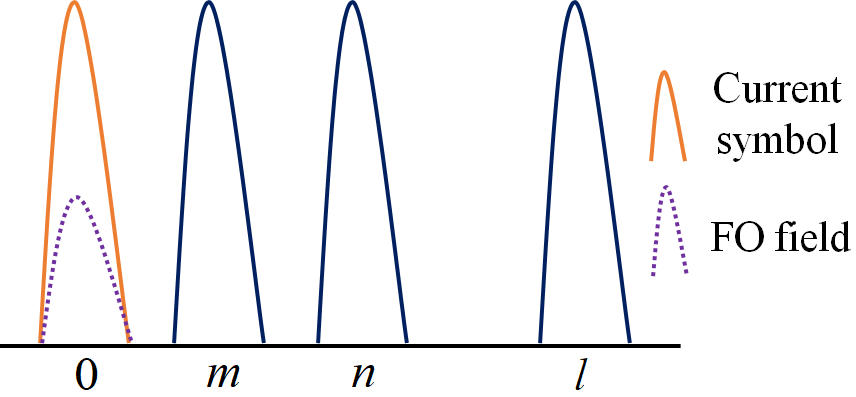}
					\par\end{flushleft}
				\begin{center}
					\hspace{0.8cm} 	
					\vspace{-0.2cm} 
					
					(a)
					\par\end{center}%
			\end{minipage}\hspace{2cm}%
			\begin{minipage}[t]{0.48\columnwidth}%
				\begin{flushright}
					\includegraphics[width=0.4\paperwidth,height=0.1\paperheight]{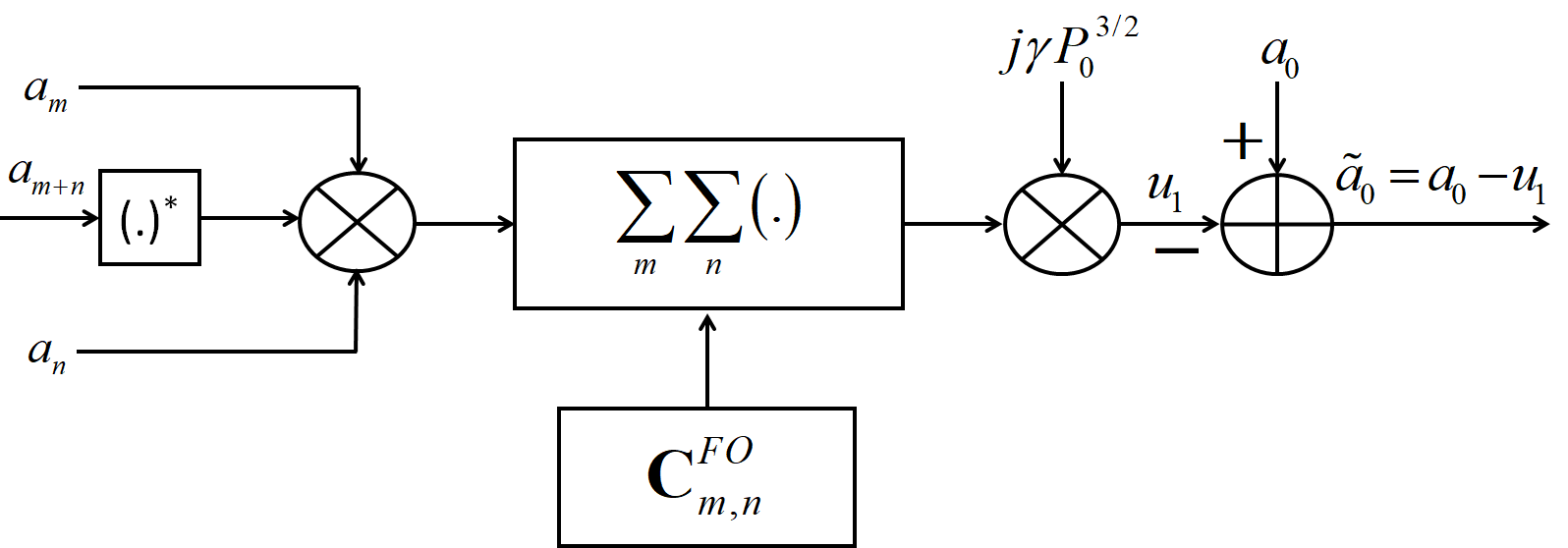}
					\par\end{flushright}
				\begin{center}
					\hspace{3.5cm} 
					\vspace{-0.6cm} 
					(b)
					\par\end{center}%
			\end{minipage}
			\par\end{center}%
	\end{minipage}
	\centering{}\caption{The FO-PB-NLC technique. (a) triplet pulses involving in the FO distortion field calculation and (b) the block diagram of the FO-PB-NLC technique. }
\end{figure*}

\begin{multline}
\label{eqn15}
\textbf{C}_{m,n}^{FO}=\intop_{0}^{L}\frac{\exp(-\alpha z)}{\sqrt{1+2j\beta_{2}z/\tau^{2}+3(\beta_{2}z/\tau^{2})^{2}}}\\
\times\exp\left(-3\frac{m{\it nT}^{2}}{\tau^{2}(1+3j\beta_{2}z/\tau^{2})}\right.
{\displaystyle \left.-\frac{\left(m-n\right)^{2}T^{2}}{\tau^{2}[1+2j\beta_{2}z/\tau^{2}+3(\beta_{2}z/\tau^{2})^{2}]}\right)dz.}
\end{multline}
   
In the FO-PB-NLC technique, the perturbation coefficient matrix $\textbf{C}_{m,n}^{FO}$ is calculated offline and stored in a look-up table. The basic idea of the predistortion technique is to calculate the FO nonlinear distortion field $u_{1}$ firstly using (\ref{eqn14}) and then to subtract it from the symbol under consideration (i.e., the symbol at the zeroth index) $a_{0}$ to generate the predistorted symbol $\tilde{a_{0}}$, as shown in Fig. 2(b). 

\section{Theory of the Second-Order Perturbation-Based Predistortion Technique}

In this section, we discuss the theory of the SO perturbative correction to the nonlinear distortion field in the context of a single-channel and single-polarization transmission system, which is the foundation of the SO-PB-NLC technique. 

The differential equation governing the SO distortion field is obtained by substituting $k'=2$ in (\ref{eqn8}), which can be represented as: 

\begin{equation}
\label{eqn16}
\frac{\partial}{\partial z}u_{2}(z,t)=\underset{\textrm{Linear part}}{\underbrace{-j\frac{\beta_{2}}{2}\frac{\partial^{2}}{\partial t^{2}}u_{2}(z,t)}}+\underset{\textrm{Nonlinear part}}{\underbrace{\left(\begin{array}{c}
		j\underset{\textrm{Term 1}}{\underbrace{2\left|u_{0}(z,t)\right|^{2}\tilde{u}_{1}(z,t)\exp(-\alpha z)}}\\
		+j\underset{\textrm{Term 2}}{\underbrace{u_{0}^{2}(z,t)\tilde{u}_{1}^{*}(z,t)\exp(-\alpha z)}}
		\end{array}\right)}},
\end{equation}
where $\tilde{u}_{1}$ is the FO field distorted by CD in the incremental length of $z$ while evolving along the optical fiber. It is important to mention that the dispersed FO ghost pulse is considered in the calculation of the SO distortion field. 

The (\ref{eqn16}) represents the evolution of the SO distortion field along the dispersive and nonlinear optical fiber channel. In (\ref{eqn16}), the linear part represents the dispersion effect on $u_{2}$ when it propagates in the fiber. Term 1 of nonlinear part represents the intra-channel cross-phase modulation between the zeroth-order and the FO distortion fields, whereas Term 2 is the intra-channel four-wave mixing term. 

In contrast to the triplet pulses-induced nonlinear ghost pulse generation in the FO perturbation theory, the SO distortion field $u_{2}$ is generated by nonlinear interaction between quintuplet pulses. The triplet pulses located at arbitrary time indices $m,\,n,\,$ and $l$ generate the FO ghost pulse at the time index $m+n-l$. That will further interact with linearly dispersed pulses at $k$ and $p=m+n-l+k$, and generate SO distortion at the zeroth index. Accordingly, in congruous with the recursive nature of the RP-based approximation, the SO ghost pulse is generated by the nonlinear interaction between the FO ghost pulse and two other linearly dispersed pulses. 

\begin{table*}[t]
	\setlength{\jot}{-3pt}
	\begin{multline}
	\label{eqn17}
	\textbf{C}_{m,n,l,k}^{SO,\,Term\,1}=-\tau^{4}\intop_{0}^{L}\intop_{0}^{z}\frac{\exp(-\alpha(z+s))}{\sqrt{-\overline{A}(z,s)\overline{B}(s)}}\exp\left\{ \frac{2T^{2}}{\tau^{2}\overline{A}(z,s)\overline{B}(s)}\right.\left[\vphantom{\frac{1}{4}s^{2}z^{2}\beta_{2}^{4}L_{m,n,l,k}}\overline{C}_{m,n,l,k}\tau^{8}\right.\\
	+j\beta_{2}(\overline{D}_{m,n,l,k}s-\overline{E}_{m,n,l,k}z)\tau^{6}+\frac{3}{2}\beta_{2}^{2}(\overline{F}_{m,n,l,k}s^{2}-\overline{G}_{m,n,l,k}sz+\overline{H}_{m,n,l,k}z^{2})\tau^{4}\\
	+\frac{3}{2}j\beta_{2}^{3}sz(\overline{I}_{m,n,l,k}s-\overline{J}_{m,n,l,k}z)\tau^{2}\left.\left.+\frac{1}{4}s^{2}z^{2}\beta_{2}^{4}\overline{K}_{m,n,l,k}\right]\vphantom{\frac{2T^{2}}{\tau\hat{A}^{2}(z,s)\hat{C}(s)}}\right\} \,ds\,dz,
	\end{multline}
	where\setlength{\jot}{1pt}
	\begin{flalign}
	& \overline{A}(z,s)=\left({\displaystyle j\tau^{6}-3\beta_{2}\left(s+2/3z\right)\tau^{4}-6j\beta_{2}^{2}\left(s-7/6z\right)z\tau^{2}}-5sz^{2}\beta_{2}^{3}\right),\\
	& {\displaystyle \overline{B}(s)=\left(j\tau^{2}+\beta_{2}s\right)},\\
	& \overline{C}_{m,n,l,k}={\displaystyle \left(7/4l^{2}+\left(-k/2-2m-2n\right)l+3/4m^{2}+\left(k/2+n\right)m\right.}\nonumber \\
	& \left.\hspace{6cm}+1/2nk+1/2k^{2}+3/4n^{2}\right),\\
	& {\displaystyle \overline{D}_{m,n,l,k}=\left(-l^{2}+\left(-k+m/2+n/2\right)l+1/2m^{2}+\left(k-n/2\right)m+nk+1/2n^{2}+k^{2}\right),}\\
	& \overline{E}_{m,n,l,k}=\left({\displaystyle -5l^{2}+\left(7/2k+6(m+n)\right)l-2m^{2}+7/2\left(-3/7k-6/7n\right)m}\right.\nonumber \\
	& \left.\hspace{10cm}-1/2\left(3k+4n\right)n\right),\\
	& \overline{F}_{m,n,l,k}=\left({\displaystyle 1/2l^{2}-\left(m+n+k\right)l+1/2m^{2}+\left(n+k\right)m+nk+1/2n^{2}+k^{2}}\right),\\
	& \overline{G}_{m,n,l,k}=\left({\displaystyle -10/3l^{2}+\left(10/3k+4(m+n)\right)l-2/3m^{2}+10/3\left(-3/5k-n\right)m}\right.\nonumber \\
	& \left.\hspace{10cm}-2/3\left(3k+n\right)n\right),\\
	& {\displaystyle \overline{H}_{m,n,l,k}=\left(7/2l^{2}+7/2\left(-4/3m-4/3n\right)l+2nm+13/6(m+n^{2})\right),}\\
	& {\displaystyle \overline{I}_{m,n,l,k}=\left(k\left(l-m-n\right)\right),}\\
	& {\displaystyle \overline{J}_{m,n,l,k}=\left(10/3l^{2}+1/3\left(-13(m+n)\right)l+m^{2}+11/3nm+n^{2}\right),}\\
	& \overline{K}_{m,n,l,k}=\left({\displaystyle \left(l-m-n\right)^{2}}\right).
	\end{flalign}
	\rule[0.5ex]{1\textwidth}{0.5pt}
\end{table*}  

For simplicity of analysis, we consider Term 1 and Term 2 of the nonlinear part in (\ref{eqn16}) separately, and finally, combine them.     

	\textit{When Term 1 of (\ref{eqn16}) is considered, the coefficient of nonlinear interaction between five input Gaussian pulses $\sqrt{P_{0}}a_{m/n/l/k/p}\exp(-(t-T_{m/n/l/k/p})^{2}/2\tau^{2})$ at $T_{m},\,T_{n},\,T_{l},\,T_{k},\,T_{p}$ and substituting the phase-matching condition $p=m+n-l+k$ and $t=0$ (symbol rate operation), can be expressed as (\ref{eqn17}).}
	

Next, we consider Term 2 of the nonlinear part in (\ref{eqn16}) to carry out the analysis.   

		\textit{When Term 2 of (\ref{eqn16}) is considered, the coefficient of nonlinear interaction between five input Gaussian pulses $\sqrt{P_{0}}a_{m/n/l/k/p}\exp(-(t-T_{m/n/l/k/p})^{2}/2\tau^{2})$ at $T_{m},\,T_{n},\,T_{l},\,T_{k},\,T_{p}$ and substituting the phase-matching condition $p=m+n-l+k$ and $t=0$, can be expressed as (\ref{eqn29}).}
\begin{table*}[t]
	\setlength{\jot}{1pt}
	\begin{multline}
	\label{eqn29}
	\textbf{C}_{m,n,l,k}^{SO,\,Term\,2}=\sqrt{3}\tau^{4}\intop_{0}^{L}\intop_{0}^{z}\frac{\sqrt{\widehat{A}(z,s)}\exp(-\alpha(z+s))}{\sqrt{\widehat{B}(z,s)\widehat{C}(z)}\left(\sqrt{-\overline{B}(s)\widehat{D}(z,s)}\right)^{*}}\exp\left\{ \frac{-jT^{2}}{2\tau^{2}\widehat{B}(z,s)\widehat{E}(z)\widehat{F}(s)}\right.\\
	\left[\widehat{G}_{m,n,l,k}\tau^{8}+2j\beta_{2}(\widehat{H}_{m,n,l,k}z+\widehat{I}_{m,n,l,k}s)\tau^{6}-3\beta_{2}^{2}(\widehat{J}_{m,n,l,k}z^{2}-\widehat{K}_{m,n,l,k}sz+\widehat{L}_{m,n,l,k}s^{2})\tau^{4}\right.\\
	-2j\beta_{2}^{3}sz(\widehat{M}_{m,n,l,k}z+\widehat{N}_{m,n,l,k}s)\tau^{2}\left.\left.-s^{2}z^{2}\beta_{2}^{4}\widehat{O}_{m,n,l,k}\right]\vphantom{\frac{-jT^{2}}{2\tau^{2}\widehat{B}(z,s)\widehat{E}(z)\widehat{F}(s)}}\right\} \,ds\,dz,
	\end{multline}
	where\setlength{\jot}{-1pt}
	\begin{flalign}
	& \widehat{A}(z,s)=\left({\displaystyle {\it j\tau^{4}+js}z\beta_{2}^{2}+3\tau^{2}\beta_{2}\left(s-z\right)}\right),\\
	& {\displaystyle \widehat{B}(z,s)=\left({\displaystyle \tau^{4}-3j\left(s-7/3z\right)\beta_{2}\tau^{2}+5sz\beta_{2}^{2}}\right)},\\
	& {\displaystyle \widehat{C}(z)=\left({\displaystyle j\tau^{2}+\beta_{2}z}\right),}\\
	& \widehat{D}(z,s)={\displaystyle \left({\displaystyle \tau^{4}+sz\beta_{2}^{2}+{\it j3}\tau^{2}\beta_{2}\left(s-z\right)}\right),}\\
	& \widehat{E}(z)=\left({\displaystyle j\beta_{2}z-\tau^{2}}\right),\\
	& \widehat{F}(s)=\left({\displaystyle {\displaystyle j\tau^{2}-\beta_{2}s}}\right),\\
	& \widehat{G}_{m,n,l,k}=\left({\displaystyle -\left(m+n-l+k\right)^{2}-k^{2}-6l^{2}+6\left(m+n\right)l-2(m^{2}+mn+n^{2}})\right),\\
	& \widehat{H}_{m,n,l,k}=\left(-3\left(m+n-l+k\right)^{2}+\left(k+2(2l-m-n)\right)\left(m+n-l+k\right)-3k^{2}\right.\nonumber \\
	& \left.{\displaystyle \hspace{4cm}+\left(2(2l-m-n)\right)k-2l^{2}}+2\left(m+n\right)l-2(m^{2}-mn+n^{2})\vphantom{-3\left(m+n-l+k\right)^{2}}\right),\\
	& \widehat{I}_{m,n,l,k}=\left({\displaystyle \left(m+n-l+k\right)^{2}+k^{2}-3\left(-n+l\right)\left(l-m\right)}\right),\\
	& {\displaystyle \widehat{J}_{m,n,l,k}={\displaystyle \left(\left(m+n-l+k\right)^{2}+\left(2k-8/3l+4/3m+4/3n\right)\left(m+n-l+k\right)\right.}}\nonumber \\
	& \left.\hspace{2cm}+k^{2}-4/3\left(2l-m-n\right)k+10/3l^{2}-10/3\left(m+n\right)l+2m^{2}+2n^{2}-2/3mn\vphantom{\left(m+n-l+k\right)^{2}}\right),\\
	& {\displaystyle \widehat{K}_{m,n,l,k}=\left(4/3\left(m+n-l+k\right)^{2}+4/3\left(k-2l+m+n\right)\left(m+n-l+k\right)\right.}\nonumber \\
	& \hspace{9cm}\left.+4/3\left(n-l+k\right)\left(m-l+k\right)\vphantom{4/3\left(m+n-l+k\right)^{2}}\right),\\
	& {\displaystyle \widehat{L}_{m,n,l,k}=\left({\displaystyle \left(m+n-l+k\right)^{2}+k^{2}}\right),}\\
	& \widehat{M}_{m,n,l,k}={\displaystyle \left(\left(m+n-l+k\right)^{2}+2\left(k-2l+m+n\right)\left(m+n-l+k\right)\right.}\nonumber \\
	& \left.\hspace{5cm}+k^{2}-2\left(2l-m-n\right)k+5\left(-n+l\right)\left(l-m\right)\vphantom{\left(m+n-l+k\right)^{2}}\right),\\
	& {\displaystyle \widehat{N}_{m,n,l,k}=\left({\displaystyle \left(m+n-l+k\right)^{2}-3\left(m+n-l+k\right)k+k^{2}}\right),}\\
	& {\displaystyle \widehat{O}_{m,n,l,k}=\left({\displaystyle {\displaystyle \left(m+n-l+2k\right)^{2}}}\right).}
	\end{flalign}
	\rule[0.5ex]{1\textwidth}{0.5pt}
\end{table*}


 The 4-dimensional (D) matrices $\textbf{C}_{m,n,l,k}^{SO,\,Term\,1}$ and $\textbf{C}_{m,n,l,k}^{SO,\,Term\,2}$ represent the coefficients of nonlinear interaction between the quintuplet pulses.       

	\textit{When considering both Term 1 and Term 2 of (\ref{eqn16}), the five input Gaussian pulses $\sqrt{P_{0}}a_{m/n/l/k/p}\exp(-(t-T_{m/n/l/k/p})^{2}/2\tau^{2})$ at five timings $T_{m},\,T_{n},\,T_{l},\,T_{k},\,$ and $T_{p}$, where $p=m+n-l+k$ is the phase-matching condition, generate the SO distortion pulse at zeroth index (with symbol rate operation assumption), which can be represented as:} 
	
	\begin{equation}
	\label{eqn45}
	u_{2}(L,t)=\gamma^{2}P_{0}^{5/2}\sum_{m}\sum_{n}\sum_{l}\sum_{k}
	\left[\vphantom{\textbf{C}_{m,n,l,k}^{SO,\,Term\,2}}2a_{m}a_{l}^{*}a_{n}a_{k}a_{m+n-l+k}^{*}\textbf{C}_{m,n,l,k}^{SO,\,Term\,1}\right.
	\left.+a_{m}^{*}a_{l}a_{n}^{*}a_{k}a_{m+n-l+k}\textbf{C}_{m,n,l,k}^{SO,\,Term\,2}\right],
	\end{equation}
\textit{where} $\textbf{C}_{m,n,l,k}^{SO,\,Term\,1}$ \textit{and} $\textbf{C}_{m,n,l,k}^{SO,\,Term\,2}$ \textit{are given by (\ref{eqn17}) and (\ref{eqn29}), respectively.}

The SO predistortion technique using (\ref{eqn45}) is also based on the same simplifying assumptions considered for the FO-PB-NLC technique, such as the full electronic compensation of the CD effect at the receiver and the Gaussian shape assumption for the input pulse shape. The SO nonlinearity coefficient matrices ${\textbf{C}}_{m,n,k}^{SO,\,Term\,1}$ and ${\textbf{C}}_{m,n,k}^{SO,\,Term\,2}$ are calculated offline and stored in look-up tables. Then, the SO distortion field ${u}_{2}$ is calculated using (\ref{eqn45}) and subtracted from the zeroth index symbol $a_{0}$ to generate the predistorted symbol. 

The SO-PB-NLC technique uses (\ref{eqn45}) to calculate the SO nonlinear distortion field. It considers the nonlinear interaction of the quintuplet pulses located at all possible arbitrary time indices. However, (\ref{eqn45}) is practically unrealizable when the possible combinations of the dispersed symbols with symbol indices $m,\,n,\,l$, and $k$ approach infinity. Consequently, we put a cap on the maximum number of the perturbation terms in the calculation of (\ref{eqn45}) by introducing a truncation threshold for the 4-D nonlinear coefficient matrices $\textbf{C}_{m,n,l,k}^{SO,\,Term\,1}$ and $\textbf{C}_{m,n,l,k}^{SO,\,Term\,2}$. The truncation threshold can be defined as the threshold at which the magnitude of $\textbf{C}_{m,n,l,k}^{SO,\,Term\,1/Term\,2}$ is less than the maximum magnitude $C_{0,0,0,0}^{SO,\,Term\,1/Term\,2}$ by a factor $\mu$, i.e., $20\textrm{\,log}_{10}\left(\left|\textbf{C}_{m,n,l,k}^{SO,\,Term\,1/Term\,2}\right|/\left|C_{0,0,0,0}^{SO,\,Term\,1/Term\,2}\right|\right)<\mu$. 

In the proposed SO-PB-NLC technique, we adopt a quantization method proposed in \cite{ZTao2013} to reduce the computational complexity further. It is important to mention that the nonlinearity coefficients in ${\textbf{C}}_{m,n,l,k}^{SO,\,Term\,1}$ and ${\textbf{C}}_{m,n,l,k}^{SO,\,Term\,2}$ are very similar, in particular for those with large indexes. Based on this fact, we ignore the coefficient difference of $\pm0.5$, which will dramatically reduce the number of nonlinearity coefficients satisfying the thresholding condition. That will significantly reduce the implementation complexity of the SO-PB-NLC technique.         

\section{Simulation Parameters and Results}

At the transmitter, after the root-raised cosine (RRC) pulse shaping, the predistorted signal is up-converted to the optical domain and transmitted over the long-haul optical fiber link. The simulation parameters used for the study are listed in Table 1. The modulation format used is 16-QAM. The symbol rate considered is 32 Gbaud. The amplified spontaneous emission (ASE) noise of EDFA is added to the signal after each fiber span to capture the nonlinear interaction between the signal and the ASE noise \cite{Irukulapati2014}.   

\begin{table}[!t]
	\renewcommand{\arraystretch}{1.2}
	\caption{Simulation Parameters.}
	\centering
	\begin{tabular}{|c|c|}
		\hline
		Parameter & Value\\
		\hline
		\hline
		RRC filter roll-off factor & $0.1$ \\
		\hline
		$\mu$ & $-40$ dB \\
		\hline
		Fiber span length & $80$ km\\
		\hline
		$\alpha$ & $0.2$ dB/km\\
		\hline 
		$\beta_{2}$ & $-20.47$  $\textrm{p\ensuremath{\textrm{s}^{2}}}$/km\\
		\hline 
		$\gamma$ & $1.22$  (1/W)/km\\
		\hline 
		$\textrm{Polarization mode dispersion coefficient}$ & $0.1$  $\textrm{ps/\ensuremath{\sqrt{\textrm{km}}}}$\\
		\hline 
		$\textrm{Noise figure of EDFA }$ & $5.5$  dB\\
		\hline  
	\end{tabular}
\end{table}

\subsection{Simulation Results}  

Fig. 3 shows the bit error rate (BER) as a function of the launch power for the SO-PB-NLC, FO-PB-NLC, and EDC techniques for a single-channel and single-polarization optical transmission system. The BER performance of the benchmark DBP technique implemented with one step per span is also included for comparison. The transmission distance considered is 2800 km. We observe that the BER performance of the SO-PB-NLC technique is significantly better than that of the FO-PB-NLC and EDC techniques. Another observation is that the BER performance of DBP is higher than that of the proposed SO-PB-NLC technique. That is because the DBP is a numerical method that uses the SSFM, and so it compensates for the nonlinearity effects span-by-span. On the other hand, the PB-NLC techniques use an analytical approximation for the solution of the NLSE with the assumption that the fiber link has only one span. That is a general assumption considered in the design of the PB-NLC techniques. It is important to mention that the single span assumption of the PB-NLC techniques allows the compensation of the nonlinearity effect in a single computation step, thus reducing the computational effort required.

\begin{figure*}[t]
	\centering{}%
	\noindent\begin{minipage}[t]{0.45\columnwidth}%
		\begin{center}
			\includegraphics[width=0.95\columnwidth,height=0.19\paperheight]{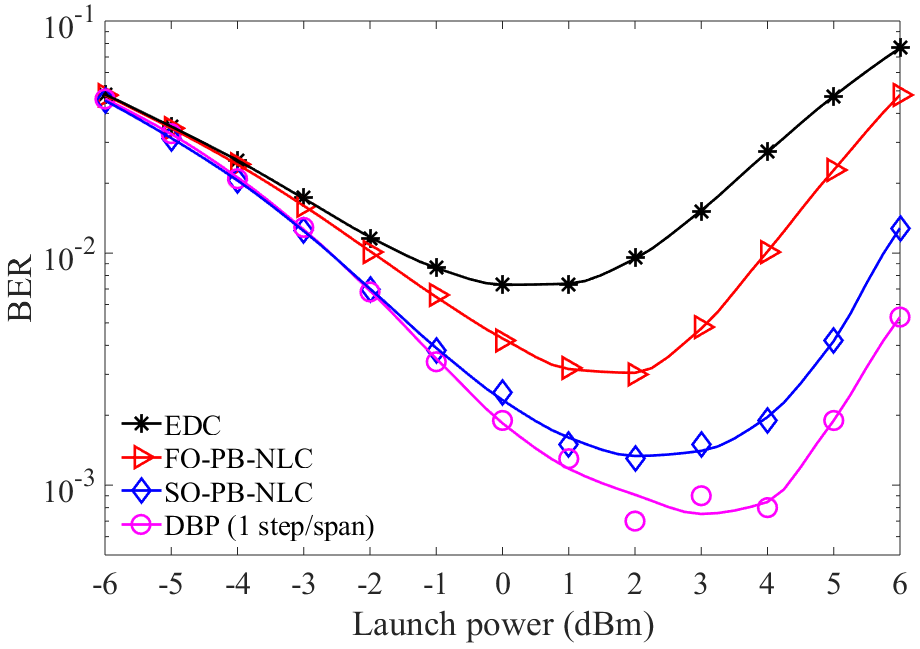}\caption{BER vs. launch power.}
			\par\end{center}%
	\end{minipage}\hspace{0.4cm}%
	\noindent\begin{minipage}[t]{0.45\columnwidth}%
		\begin{center}
			\noindent\begin{minipage}[t]{1\columnwidth}%
				\begin{center}
					\includegraphics[width=0.95\columnwidth,height=0.19\paperheight]{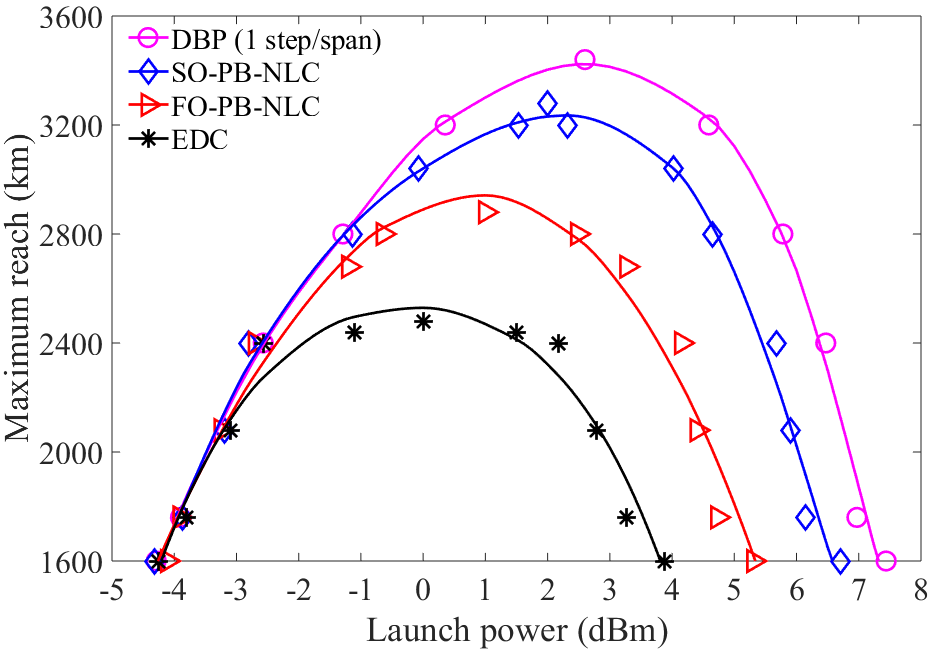}
					\caption{Maximum reach as a function of the launch power.}
					\par\end{center}%
			\end{minipage}
			\par\end{center}%
	\end{minipage}
\end{figure*}
    
Fig. 4 presents the plot of the maximum system reach for the considered techniques at 7\% overhead (OH) hard-decision (HD) forward error correction (FEC) limit with a BER value of $4.3\times10^{-3}$. It is observed that the maximum transmission reach for DBP, SO-PB-NLC, FO-PB-NLC, and EDC is 3440 km, 3280 km, 2880 km, and 2480 km, respectively. These results indicate that the SO-PB-NLC technique provides an extended transmission reach by 32.2\% and 14\% when compared to EDC and the FO-PB-NLC techniques, respectively. It can also be inferred from Fig. 4 that the nonlinearity threshold of the DBP, SO-PB-NLC, and FO-PB-NLC techniques is improved by 6.3 dB, 5.3 dB, and 3.6 dB, respectively, when compared to the EDC technique at a transmission distance of 2480 km (i.e., the maximum reach for the EDC technique). The nonlinearity threshold is defined as the value of the launch power at which the BER performance crosses the FEC limit for a given transmission distance \cite{LBDu2012}. It is also observed that the nonlinearity threshold of the SO-PB-NLC technique is improved by 1.7 dB when compared to the FO-PB-NLC technique.  

\renewcommand{\arraystretch}{1.3}

\vspace{-0.5cm}
\section{Conclusion}

In this paper, we have proposed to extend the FO-PB-NLC technique to the SO, referred to as the SO-PB-NLC technique. We have derived the SO nonlinear distortion field for a single-channel and single-polarization system with a Gaussian shape assumption for the input pulse. Through simulations, we have shown that the NLC performance of the SO-PB-NLC technique is better than that of the FO-PB-NLC technique. 


%

\end{document}